\documentclass[sigconf]{acmart}
\def\BibTeX{{\rm B\kern-.05em{\sc i\kern-.025em b}\kern-.08emT\kern-.1667em\lower.7ex\hbox{E}\kern-.125emX}}

\setcopyright{rightsretained}
\copyrightyear{2019}
\acmYear{2019}
\acmConference[FDG '19]{The Fourteenth International Conference on the Foundations of Digital Games}{August 26--30, 2019}{San Luis Obispo, CA, USA}
\acmDOI{10.1145/3337722.3341836}
\acmISBN{978-1-4503-7217-6/19/08}

\begin{document}

%
\title{Rhythm Dungeon: A Blockchain-based Music Roguelike Game}

%

\author{Tengfei Wang}
\affiliation{%
  \institution{The Chinese University of Hong Kong, Shenzhen}
  \streetaddress{2001 Longxiang Ave}
  \city{Shenzhen, China}
}
\email{tengfeiwang@link.cuhk.edu.cn}

\author{Shuyi Zhang}
\affiliation{%
  \institution{White Matrix Inc.}
  \city{Nanjing, China}
}
\email{shuyizhang@matrixdapp.com}

\author{Xiao Wu}
\affiliation{%
  \institution{White Matrix Inc.}
  \city{Nanjing, China}
}
\email{ling@matrixdapp.com}

\author{Wei Cai}
\authornote{corresponding author}
\affiliation{%
  \institution{The Chinese University of Hong Kong, Shenzhen}
  \streetaddress{2001 Longxiang Ave}
  \city{Shenzhen, China}
}
\email{caiwei@cuhk.edu.cn}

%

%
\begin{abstract}
Rhythm Dungeon is a rhythm game which leverages the blockchain as a shared open database. During the gaming session, the player explores a roguelike dungeon by inputting specific sequences in time to music rhythm. By integrating smart contract to the game program, the enemies through the venture are generated from other games which share the identical blockchain. On the other hand, the player may upload their characters at the end of their journey, so that their own character may appear in other games and make an influence. Rhythm Dungeon is designed and implemented to show the potential of decentralized gaming experience, which utilizes the blockchain to provide asynchronous interactions among massive players.
\end{abstract}

%
%



%

%

%
\maketitle

\section{Introduction}

By leveraging the unique features of blockchain technology \cite{Nofer2017}, decentralized ledger, e.g. BitCoin \cite{bitcoin}, becomes reality. On this basis, Ethereum \cite{ethereum} further extends the application of blockchain into the domain of decentralized applications (dApps) \cite{WeiCaiWEHFL2018} by introducing the concept of smart contract \cite{smartcontractlogistics}. Since 2017, tremendous attentions have been attracted to develop a variety of applications over blockchain platforms to enjoy the benefits of decentralization: immutable, transparent and auditable data and functionalities. Blockchain game \cite{MinTWGC} is one of the most active categories of dApps in the market. CrytopKitties\footnote{https://www.cryptokitties.co/}, the most popular blockchain game, once dominated the traffic in Ethereum network due to its popularity. It leverages the benefits of immutable virtual asset tokens and transparent game rules to improve players' gaming experience. In fact, blockchain has the potential to redefine the relationship model between game operators and players: the game data will be possessed by the players rather than the game operators. Among the benefits blockchain brought to the games, the interoperability of game data attracts the most attention. For instance, the CryptoCuddles\footnote{https://cryptocuddles.com/} allows the players to use their own kitties purchased at CryptoKitiies to battle. However, the state-of-the-art blockchain games can only reuse the existing tokens available in a single direction, which means that the data modification in new games can not make effects on the token they inherit. This implies that, the relationship between the new and the existing games are principal and subordinate, rather than equal interoperability. In this demonstration\footnote{Demo video available at: http://gaim.sse.cuhk.edu.cn/rhythm-dungeon/}, we design and implement Rhythm Dungeon, a music roguelike game to interact with Last Trip and Adam's Adventure (AA), another two blockchain games we developed to demonstrate the concept of interoperability \cite{CaiW2019}.

\section{System Framework}

\begin{figure}[htbp]
\centering\includegraphics[width = 3.35in]{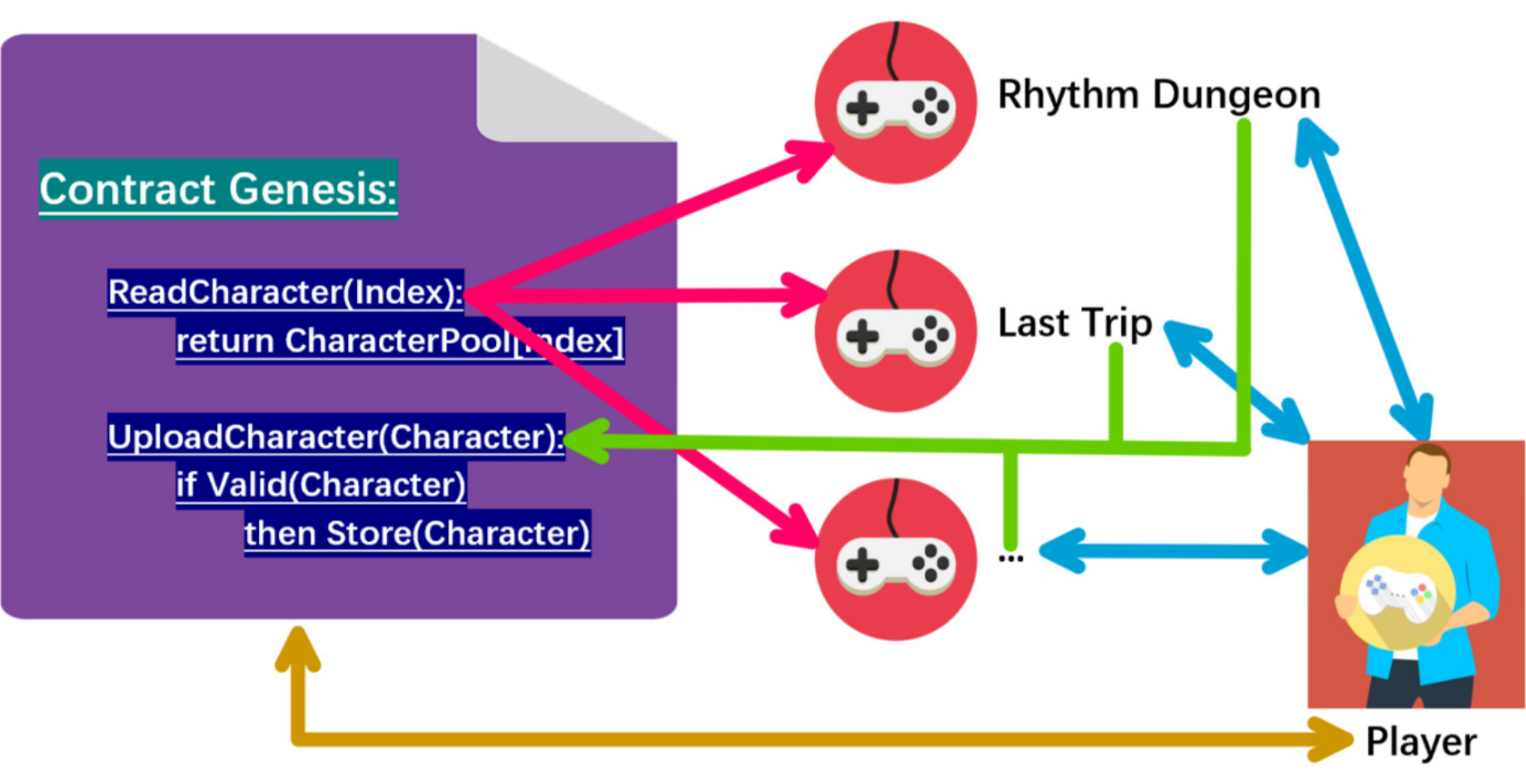}
\caption{System Framework}
\label{fig:framework}
\end{figure}

\begin{figure*}[htbp]
\centering\includegraphics[width = 7in]{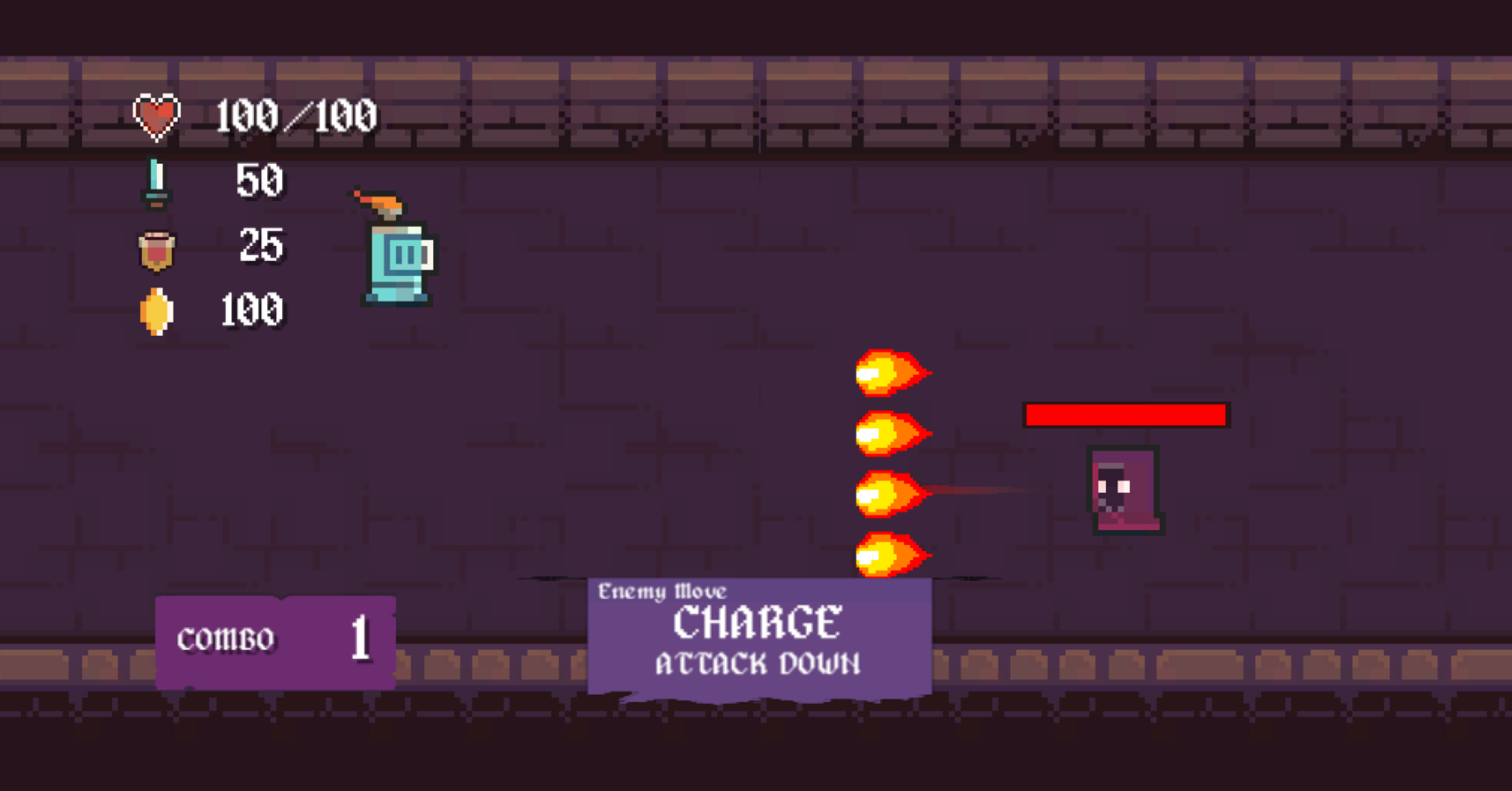}
\caption{Screenshot of Rhythm Dungeon: The player's character (left) is fighting against the enemy. }
\label{fig:screenshot}
\end{figure*}

In order to enable data interoperability, we implement packet Genesis\footnote{https://github.com/SFENKSLab/ProjectManis/tree/master/Packages/ProjectGenesis}, an open source smart contract package that hosts interoperable data for multiple games in the blockchain. Due to the length limit, we would like to skip the implementation details but only introduce the interfaces for our demo. As illustrated in Figure \ref{fig:framework}, the Genesis smart contract provides two public functions, ReadCharacter and UploadCharacter, to facilitate interoperability. The games developed with Genesis framework are able to fetch characters from Genesis via the ReadCharacter function. Also, after the function UploadCharacter is called with a character as the parameter, the contract stores the character if it is valid after checking. Finally, when players play the games that have a connection with Genesis, they can experience the features of the blockchain and contribute contents to the contract by playing games. As an open source framework, all players have free access to the content of the contract and can read data that it has stored.

Based on the framework, we have designed and implemented two blockchain games: Last Trip and Adam's Adventure (AA). Last Trip is a storybook, in which the players perform different actions to improve their characters' attributes to win the battles. The player can upload his character to the blockchain after the game is over. AA is a multi-player Dungeons and Dragons (D\&D) novel-adventure game. In this game, the players will experience three modes: Battle of Adventure, Battle of Dark Lord, and Battle of Blood Moon. In the Battle of Adventure mode, massive players will create characters to adventure in the Dungeon, conquer demons to empower their characters on their own. All growth in these characters will eventually be accumulated into Adam, a shared character stored in the blockchain. In the battle of Dark Lord, the player will use the shared Adam and one random character summoned from the Last Trip to combat the Dark Lord, the evil king created by the system. Each 30 times Dark Lord is defeated in every chain, the battle of Blood Moon can be triggered. The player will use the Adam in their corresponding chain to conquer the Adams in other chains.

The Rhythm Dungeon will interact with the characters uploaded by the games above. Therefore, the players in this game may interact with past players in Last Trip and Adam's Adventure. Meanwhile, the newly created characters in Rhythm Dungeon will appear in Last Trip and Adam's Adventure as well. With this approach, the players of the tame will influence future adventures and contribute their content to make the stories in these games continuously increasing.

\section{Gameplay}

The fundamental gameplay of Rhythm Dungeon is a mixture of the music game and the role-playing game (RPG) with roguelike elements. Each round the player starts with a new basic character. During the play, the player encounters enemies that are randomly generated or fetched from the smart contract Genesis, according to the level of the player's character. The player has to press buttons in sync with the drum beats to fight against enemies. Every bar contains four beats and the player has to press four buttons to deploy one action (including attacking, dodging and charging). The player can take one action every two bars. So, it is a kind of turn-based input mechanism, just like the famous PSP game, Patapon. A screenshot of Rhythm Dungeon is illustrated in Figure \ref{fig:screenshot}.

The role-playing elements of this game are quite straightforward. Player kills enemies to earn experience points to level up. When the player upgrades, he or she collects skill points to increase the character's strength, armor, luck (related to the chance of a critical hit) and max health. The career of the character is determined by the characteristic of the attributes. Due to the different tendency of the attributes, the player can adopt different strategies for enemies of different careers to kill more enemies. We plan to design more contents for careers, which will be talked about later in the Ongoing Work section.

Regarding the roguelike characteristic of this game, each time the player starts with a new character. The character grows up when it levels up. The death of the character is permanent, which means there is no chance for the character to come back to life again. However, the player can choose to upload his or her character to the smart contract Genesis deployed on Ethereum. Then, although one cannot play that character again, there is a possibility that the character is brought back to life when other player plays the game or act in other games that fetch this character from the blockchain.

\section{Game Features}

\subsection{Asynchronous Interactions among Players}

Players can interact with each other by letting others fight their own characters or challenging characters of others. This is done through the uploading and fetching characters to and from the smart contract. The uploaded character is checked by the smart contract so that its attributes do not violate the contract's pre-defined rules to regulate the process. With this mechanism, players can interact with each other, or even with themselves, in an asynchronous manner. Note that, the smart contract is completely accessible to the public. Everyone can easily see the content of the contract and read data from it. This transparency of the decentralized process makes the whole process more reliable than a centralized server handling the whole procedure.

\subsection{Interaction with Other Games}

Rhythm Dungeon, Last Trip and AA are designed to have the equal privilege to access blockchain data through the Genesis smart contract. They all have permissions to upload, fetch, and update characters. When Rhythm Dungeon fetches a character uploaded by another game, the player is in combat with a character from a different game world. Meanwhile, the characters generated by Rhythm Dungeon will appear in the other two games as well.


\subsection{Adaptive Soundtracks}

As the players encounter different types of enemies, the background soundtracks with different tempos will be played for the particular challenge. In fact, the game controls the difficulty by choosing soundtracks with distinct tempos, which require different levels of players' reaction time. In general, quicker rhythm will be associated with stronger enemies, since the players may find it more difficult to press correct key combinations in shorter intervals.


\subsection{Personalized Characters}\label{sec:personalized}

To create personalized characters for different players is a very important procedure when the system accepts the submissions of characters. In Rhythm Dungeon, the system keeps tracking the records of all the mistakes made by the players during the gaming session. When a character is uploaded, the blockchain will mark the particular character's weakness, which will be determined by the type of most frequent mistakes the players have made. With this approach, we are able to personalize the characters from the players: the other players can deliver extra damages by attacking the particular character's weak point during future battles.





\section{Ongoing Work}

The Rhythm Dungeon game is yet completed. First, the only difference among distinct character professions in current demo is the distribution of the character's attributes. In our ongoing work, special actions and unique passive skills of different professions will be implemented to enrich the gamers' experience. Second, the way we generate soundtracks should be modified to reflect the importance of character personality. Similar to the characters' weakness feature described in Section \ref{sec:personalized}, the uploaded characters' strength may affect the composition of soundtracks as well. We are looking forward to an AI-based solution to dynamically create unique music piece from the behavior of the characters. Third, there are still security concerns in connecting blockchain and game systems \cite{TianMC}. In our current demo, players should surrender the private keys of their wallets to the game server, which poses potential loss risks of players' tokens.


\section{Conclusions}

We developed Rhythm Dungeon, a music roguelike game demo which can interact with other blockchain games via the Genesis smart contract. As a turn-based roguelike RPG game, Rhythm Dungeon features the gaming experience with drum beats. More features regarding unique character professions, dynamic behavior and attributes of bosses are currently under development.

\bibliographystyle{ACM-Reference-Format}
\bibliography{library}

\end{document}